\documentclass[11pt]{article}
\pdfoutput=1

\usepackage{amsmath,amssymb,bm,epsf,epsfig,graphicx}
\input epsf.sty
\newcommand{\Tr}{\mathop{\rm Tr}\nolimits}

\def\bra#1{\langle #1 |}
\def\ket#1{|#1 \rangle}

\def \be {\begin{eqnarray}}
\def \ee {\end{eqnarray}}
\def \bdm {\begin{displaymath}}
\def \edm {\end{displaymath}}

\def\del {\partial}
\def\0{\nonumber}

\def \BCFT{{\rm BCFT}}
\def\SFT{{\rm SFT}}
\def\f{{\bar f}}
\def\F{{\mathcal F}}
\def\G{{\mathcal G}}
\def\S{{\mathcal S}}
\def\P{{\mathcal P}}

\def \hy {\hat y}

\begin{document}
\begin{center}
 {\large \bf $\,$\\
\vskip2cm
Large BCFT moduli in open string field theory}

\vskip 1.1cm

{\large Carlo Maccaferri\footnote{Email:
maccafer at gmail.com}$^{(a)}$,  Martin Schnabl\footnote{Email:
schnabl.martin at gmail.com}$^{(b)}$} \vskip 1 cm
$^{(a)}${\it Dipartimento di Fisica, Universit\'a di Torino and INFN, Sezione di Torino\\
Via Pietro Giuria 1, I-10125 Torino, Italy}\\
\vspace{.5cm}

$^{(b)}${\it {Institute of Physics of the ASCR, v.v.i.} \\
{Na Slovance 2, 182 21 Prague 8, Czech Republic}}
\end{center}

\vspace*{6.0ex}

\centerline{\bf Abstract}
\bigskip
We use the recently constructed  solution for marginal deformations
by one of the authors, to analytically relate the BCFT modulus
($\lambda_\BCFT$) to the coefficient of the boundary marginal field
in the solution ($\lambda_\SFT$). We explicitly find that the relation
is not one to one and   the same value of $\lambda_\SFT$ corresponds to a pair of different $\lambda_{\BCFT}$'s: a  ``small'' one,  and  a ``large'' one. The BCFT moduli space is fully covered,  but   the
coefficient of the marginal field in the solution is not  a good global
coordinate on such a space.  \vfill \eject

\baselineskip=15pt


\section{Introduction and Conclusion}
Open String Field Theory (OSFT) could provide a complete non-perturbative approach to
D-brane physics, once its  quantum structure is understood. Still, at the classical level it gives  a new perspective and methodology for finding the possible conformal boundary
conditions which are consistent with a given bulk two-dimensional
CFT. In String Theory language this amounts to
the classification of  the   possible physical D-branes (stable or
not) that can be consistently placed in a given closed string
background. From the OSFT perspective  this means solving the classical equation of motion.

The aim of this paper is to provide exact non-perturbative results on the relation between BCFT marginal parameters and the corresponding parameters in OSFT solutions.
Remarkably we now know \cite{EM}, that given two generic  BCFT's (sharing  a common non-compact time-like factor) it is always possible to explicitly construct an analytic solution relating the two backgrounds, by suitably using a pair of boundary condition changing operators with known OPE. However here we would like to study an example which is not so distant to the available Siegel gauge numerical results, where the time-CFT is only excited through the identity and its descendants.  For self-local boundary marginal deformations \cite{bible}, we have an explicit analytic wedge-based solution \cite{simple-marg}, and this is the solution we wish to study in this note.

Given an open string background BCFT$_0$, there is typically a
continuous manifold  of equally consistent open string
backgrounds, connected to BCFT$_0$, forming a moduli space. Such a
moduli space is
locally spanned by the VEV of the exactly marginal boundary operators that can be switched on in BCFT$_0$.
Because of the linear structure of small fluctuations, it appears
natural to parametrize the OSFT solutions for marginal
deformations by the coefficient of their marginal field. This
quantity is typically called $\lambda_{\SFT}$
\be
\Psi_{\rm marg}=\lambda_{\SFT}\;cj(0)\;\ket0_{SL(2,R)}+\cdots  .
\ee
 On the other
hand, the physical trajectory in moduli space has a more natural
coordinate $\lambda_{\BCFT}$, or more succinctly $\lambda$, which
corresponds to the strength of the (conformal) boundary
interaction that one adds to the sigma-model action to describe
the new background
\be
S_{\BCFT_\lambda}=S_{\BCFT_0}+\lambda\int_{\del M} \!ds\, j(s).
\ee
 The relation between $\lambda_\SFT$ and $\lambda_\BCFT$
triggered a lot of discussions in the last fifteen years \cite{large, toy, senT, kurs, KL1, TT-branch, large2}, especially
because of the following  long-standing puzzle. Given an exactly marginal field
$j(z)$, a one-parameter  family of approximate solutions labeled by $\lambda_\SFT$  was found
in Siegel gauge by Sen and Zwiebach \cite{large}.
 Evidence was found that this one-parameter family
 ceases to exist at a finite value of $\lambda_\SFT$, posing the question about the ability of OSFT to cover or not the BCFT moduli space.
With the advent of the new analytic methods, beginning with \cite{Schnabl}, new powerful  tools have been developed to extract the BCFT data from a given OSFT solution.
Notably it has been found how to directly construct the boundary state corresponding
to a given solution \cite{KOZ,KMS}, using a powerful conjecture due to Ellwood \cite{Ellwood}, which relates simple gauge invariants in OSFT to closed string tadpoles in
the new open string background defined by a given solution. Using Ellwood conjecture, more recent results for the cosine deformation at the self-dual radius \cite{large2},  showed that the finite critical value at which the solutions of \cite{large} truncate, correspond  to a finite value of $\lambda_\BCFT$, close to the point where the boundary conditions become Dirichlet.  After that no further solutions are found.   Where is the missing region of the BCFT moduli space?

In this note we  propose that such an apparent drawback is simply a consequence
of the fact that $\lambda_\SFT$ does not globally parameterize
OSFT solutions for marginal deformations which, on the other hand,
exist for all physical values of $\lambda_\BCFT$.
To do so, we derive, in an explicit computable example, the precise
relation between $\lambda_{\SFT}$ and $\lambda_{\BCFT}$,
 taking advantage of the  recently constructed solution for marginal deformations \cite{simple-marg}, which is naturally defined in terms of
  $\lambda_{\BCFT}$. This allows to $calculate$  $\lambda_{\SFT}$ as
a function of $\lambda_{\BCFT}$, by simply computing the coefficient of the marginal field in the solution
\be\lambda_\SFT=\langle 0|j_1c_{-1}c_0|\Psi(\lambda_\BCFT)\rangle=f_\Psi(\lambda_\BCFT).\ee

This computation gives  a nice surprise: we find that
$\lambda_{\SFT}$, as a function of $\lambda_{\BCFT}$, starts
linearly with unit slope and then, after reaching a
maximum, it starts decreasing and it eventually relaxes to zero for large values of $\lambda_{\BCFT}$, see figure \ref{fig:figure 2}. Therefore, for a given $\lambda_\SFT$  there are typically {\it two} values of $\lambda_\BCFT$. This is our main result.

 The fine details of the function $\lambda_\SFT(\lambda)$, including the critical value of $\lambda_\BCFT$ at which $\lambda_\SFT$ has a maximum, depend on the gauge freedom in the definition of the OSFT solution, but we find that the relaxation to zero is generic in the whole gauge orbit which we analyze.
  It is amazing to realize that this is precisely the behavior that Zwiebach conjectured
 many years ago \cite{toy}, by analyzing a simple field theory model for tachyon condensation. To further confirm Zwiebach's hypothesis, we also compute the coefficient
 of the zero momentum tachyon. This time, at large $\lambda$, we find that it asymptotes to a finite positive
 value. In  a particular limit along the gauge orbit the solution localizes to the boundary of the world-sheet,  and the above finite positive value
 agrees with the tachyon coefficient of the tachyon vacuum
 solution $\Psi_{TV}=\frac1{1+K}c(1+K)Bc$ of \cite{simple}, again in accord
 with Zwiebach's picture, see figure \ref{Fig:large-tach}. It is tempting to speculate that in fact the whole string field in this limit approaches the tachyon vacuum $\Psi_{TV}$ as has been shown in the case of light-like rolling tachyon in \cite{HS}.

 Our simple calculation shows that, at least in this particular example,  OSFT does cover the full BCFT moduli
 space, but such a moduli space cannot be fully described by the coefficient of the marginal field that generates the deformation. At large BCFT modulus it is not the marginal
 field that drives the marginal flow but it is rather the whole string field with all of its higher level components.

An important question  is to what degree is this behavior generic. There are other analytic wedge-based solutions for marginal deformations with singular OPE, which can be constructed  systematically at any order in the marginal parameter \cite{ KORZ, FKP, KO,KL}. However their intrinsic perturbative nature is a major obstacle to obtain conclusive results on the issues we are discussing\footnote{See \cite{L} for recent developments in this direction.}. A non perturbative treatment of (time-independent) marginal deformations is also clearly provided by the EM solution \cite{EM}.  It is not difficult to see that for this solution we have the exact relation \be\lambda_{\rm SFT}^{\textrm\small (EM)}=\lambda_{\rm BCFT},\ee which is in fact common to
all SFT solutions which describe marginal deformations with regular OPE \cite{KORZ, martin-marg, KOS}.  The reason for this is that  the EM solution in this case describes a marginal deformation \cite{KOS}  generated by $j=\frac i{\sqrt2}\del X^0+j^{(c=25)}$, which  has regular OPE with itself  by construction. Since time is non-compact, the solution only changes the boundary conditions in the $c=25$ part of the initial BCFT.

That said, it seems plausible that the double-valued dependence on $\lambda_{\rm SFT}$  we have found is generic in cases where the solution only excites the matter primaries and descendants  generated by the repeated OPE's of the marginal field.
It would be very instructive to ``experimentally" confirm this expectation by level truncation computation in the Siegel gauge, and to identify the predicted new branch.

\section{Review of the simple marginal solution}

The solution \cite{simple-marg} can be
constructed from any self-local boundary deformation
\cite{bible}, generated by a boundary field $j(x)$ with self-OPE
given by\footnote{In \cite{simple-marg} it was further assumed
that the current was not only self-local but also chiral, in the sense of \cite{bible}, so that
it was guaranteed to be local with respect to all bulk and boundary fields.
This was a technical assumption which allowed to easily construct
the fluctuations around the new solution and to show that,
for chiral marginal deformations,
 the Hilbert spaces of the undeformed and deformed theory are isomorphic at the level of the operator algebra. This is not necessarily true for generic
self-local boundary deformations.}
\be
j(x)j(0)\sim\frac1{x^2}+(reg).
\ee
 Let us
quickly review the structure of the  solution, details can be found in \cite{simple-marg}. It is derived from
an identity-based solution, discovered years ago by
Takahashi and Tanimoto \cite{TT}, which is used here as an
elementary identity-like string field in addition to the well known
fields $K,B,c$. Calling $\Phi$ the TT solution \cite{TT}, and
defining as in \cite{id-marg}
\be
K'&\equiv& K+J= Q_{\Phi\Phi}B\equiv QB+[\Phi,B],\\
J&\equiv& [B,\Phi],
\ee
where $[\cdot,\cdot]$ is the graded commutator, the solution \cite{simple-marg} can be written as
\be
\Psi=\frac1{1+K}\Phi\frac1{1+K'}-Q\left(\frac1{1+K}\Phi\frac B{1+K'}\right).\label{sol}
\ee
In the very convenient sliver frame, obtained by mapping the UHP
(with coordinate $w$) to a semi-infinite cylinder of circumference
$2$ (with coordinate $z$) via the map \be z=\frac2\pi\arctan w,
\ee
the TT solution is defined as
\be
\Phi=\int_{-i\infty}^{i\infty}\frac{dz}{2\pi
i}\left(f(z)cj(z)+\frac12 f^2(z)c(z)\right).\label{TTsol}
\ee
All the degrees
of freedom of the function $f(z)$ are pure gauge except for its zero
mode which defines $\lambda_{\BCFT}$ through the relation
 \be
\lambda_\BCFT\equiv\lambda=\int_{-i\infty}^{i\infty}\frac{dz}{2\pi
i}f(z). \ee
In the following we will make the dependence on $\lambda\equiv\lambda_\BCFT$
manifest  by defining
\be
f(z)&\equiv&\lambda\f(z),\\
\int_{-i\infty}^{i\infty}\frac{dz}{2\pi i}\f(z)&=&1.
\ee
The current-like string field $J$ is then given by
\be
J\equiv[B,\Phi]=\int_{-i\infty}^{i\infty}\frac{dz}{2\pi i}\left(f(z)j(z)+\frac12 f^2(z)\right).
\ee
\section{$\lambda_{\SFT}$ vs $\lambda_{\BCFT}$ and the tachyon}

Expanding the solution in the Fock space basis,  the
first components  are  the zero momentum tachyon and the marginal
field
\be
\Psi=T\,c_1\ket0+\lambda_{\SFT}\,j_{-1}c_1\ket0+ \cdots .
\ee
The coefficient of the marginal field is given by
\be
\lambda_\SFT&=&\bra0c_{-1}c_0j_{1}\ket\Psi=-\Tr\left[e^{-K/2}\;c\del cj\; e^{-K/2}\frac1{1+K}\Phi\frac1{1+K+J}\right]\0\\
&=&-\int_0^\infty d\ell\,e^{-\ell}\int_0^\ell dy\,\Tr\left[e^{-K/2}\;c\del
cj \;e^{-(\ell-y+1/2)K}\Phi\,e^{-y(K+J)}\right],
\ee
while the coefficient of the zero momentum tachyon is
\be
T&=&\bra0c_{-1}c_0\ket\Psi=-\frac\pi2\Tr\left[e^{-K/2}\;c\del c\; e^{-K/2}\frac1{1+K}\Phi\frac1{1+K+J}\right]\0\\
&=&-\frac\pi2\int_0^\infty d\ell\,e^{-\ell}\int_0^\ell dy\,\Tr\left[e^{-K/2}\;c\del
c \;e^{-(\ell-y+1/2)K}\Phi\,e^{-y(K+J)}\right].
\ee
Notice that the BRST exact part of the solution (\ref{sol}), does not contribute to these coefficients, as well as to any other coefficient of $c\phi^{(h)}(0)\ket0$, where $\phi$ is a matter primary.\\

\begin{figure}[!t]
\centering
\includegraphics[]{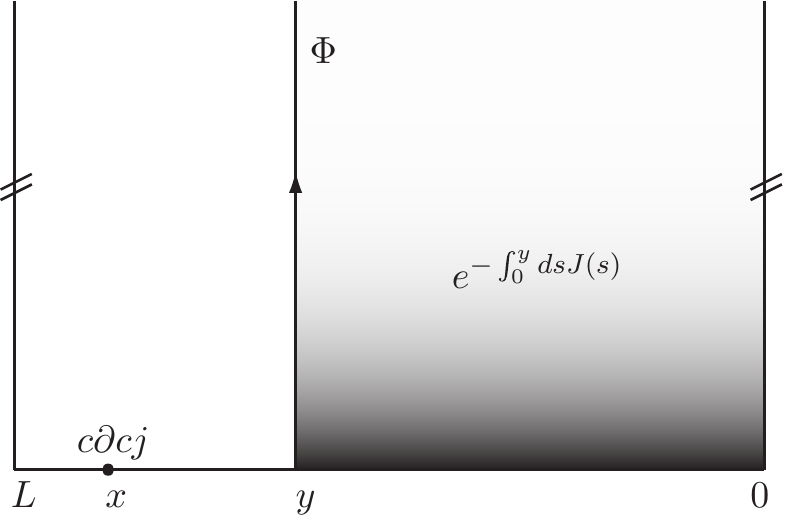}
\caption{Graphical presentation of the correlator (\ref{wedge}). The two vertical edges are identified to make a cylinder of circumference $L$. In the conventions of \cite{simple} the coordinate along the boundary increases from right to the left.
The shaded region corresponds to the insertion of the boundary interaction spread out to the bulk. On the left border of this region there is an insertion of the Takahashi-Tanimoto identity-like solution $\Phi$. }
\label{Fig:correlator}
\end{figure}

Let us start with the integrand which defines $\lambda_\SFT$
\be
\Tr\left[e^{-K/2}\;c\del cj \;e^{-(\ell-y+1/2)K}\Phi\,e^{-y(K+J)}\right]=\left\langle c\del cj(\ell+1/2)\;\Phi(y)e^{-\int_{0}^{y}ds   J(s)}   \right\rangle_{C_{\ell+1}}\label{corr}.
\ee
Here we have defined the world-sheet insertions
\be
\Phi(y)&\equiv& \int_{-i\infty}^{i\infty}\frac{dz}{2\pi i}\left(f(z) \,cj(z+y)+\frac12 f^2(z)\,c(z+y)\right),\\
J(s)&=& \int_{-i\infty}^{i\infty}\frac{dz}{2\pi i}\left(f(z) \,j(z+s)+\frac12 f^2(z)\,\right).
\ee
The correlator (\ref{corr}) is naturally defined in the cylinder coordinate  frame. General correlator of this form on a cylinder of total circumference $L$ is depicted in figure \ref{Fig:correlator}.

This correlator can be systematically computed by Wick theorem\footnote{See \cite{ZAMO}  for a general discussion.}
  from the basic current-current correlator
  \be
\langle j(z) j(w)\rangle_{C_L}=\left(\frac\pi L\right)^2\,\frac1{\sin^2\frac{\pi(z-w)}{L}},
\ee
and from the standard ghost correlator
\be
\langle c\del c(z) c(w)\rangle_{C_L}=-\left(\frac L\pi\right)^2\sin^2\frac{\pi(z-w)}{L}.
\ee
In particular, Wick theorem implies that we have
\be
\left\langle e^{-\int_{0}^{y}ds   J(s)}   \right\rangle_{C_{L}}=\exp\left[\frac12 \int_0^y\int_0^y ds_1\,ds_2 \left\langle J(s_1)J(s_2)   \right\rangle_{C_{L}}\right].
\ee
In computing the above quadruple integral (two integrals along the boundary and two vertical integrals implicit in the $J$'s) one finds out, by a mechanism analogous to  \cite{id-marg}, that
the contribution from the $f^2$-terms in $J$ precisely cancels with a delta-function contribution coming from the boundary integral of the current correlator. This leaves
us with a net result
\be
\left\langle e^{-\int_{0}^{y}ds   J(s)}   \right\rangle_{C_{L}}=e^{-\lambda^2 \G_\f(y,L)}.
\ee
The function(al) $\G_\f$ controls the exponential behavior in $\lambda_\BCFT\equiv\lambda$ and it is given by\footnote{A much quicker way to compute this correlator is to see it
as $\langle\sigma_L(y)\sigma_R(0)\rangle_{C_L}$,
where the bcc-like operators  are given by $\sigma_{L/R}(x)=e^{\mp i\lambda\chi_\f(x)}$, and use Wick theorem directly in terms of $\chi_\f$, see \cite{simple-marg} for the precise definitions.}
\be
\G_\f(y,L)&=&\int_{-i\infty}^{i\infty}\frac{d\xi}{2\pi i}\, \f\!*\!\f\,(\xi)\;\log \frac{\sin\frac{\pi(y+\xi)}{L}}{\sin\frac{\pi\xi}{L}}\label{G},\\
f\!*\!f\,(\xi)&\equiv&\int_{-i\infty}^{i\infty}\frac{dz}{2\pi i}\,f\left(z-\xi/2\right)\,f\left(z+\xi/2\right)\label{conv1}.
\ee
Then, with  standard generating function techniques, we can explicitly compute
\be
\left\langle c\del cj(x)\;\Phi(y)e^{-\int_{0}^{y}ds   J(s)}   \right\rangle_{C_{L}}=-\lambda\,{\Big(}1+\lambda^2\F_\f(x,y,L){\Big)}e^{-\lambda^2 \G_\f(y,L)}.\label{wedge}
\ee
The $\lambda^2$ contribution in front of the exponential, which we denote $\F_\f$, is given as a product of two quantities
\be
\F_\f(x,y,L)=\S_\f(x,y,L)\;\P_\f(x,y,L).
\ee
The first factor accounts for the contraction of $j(x)$,  the matter part of  the test state, with the exponential interaction and it is given by
\be
\S_\f(x,y,L)=\frac L{\pi\lambda}  \left\langle j(x)\int_0^y ds \,J(s)\right\rangle_{C_L}=\int_{-i\infty}^{i\infty}\frac{dz}{2\pi i}\f(z)\left(\cot\frac{\pi(y-x+z)}{L}+\cot\frac{\pi(x-z)}{L}\right).\label{Sf}
\ee
The second factor is responsible for the contraction between the current in $\Phi(y)$ and the exponential interaction, as well as the total ghost contribution (which gives an explicit $x$-dependence)
\be
\P_\f(x,y,L)&=&\int_{-i\infty}^{i\infty}\frac{d\xi}{2\pi i} \left[\f\!\star\!\f(\xi,x,y,L)\,\cot\frac{\pi(y-\xi)}{L}+\f\!\odot\!\f(\xi,x,y,L)\cos\frac{\pi\xi}{L}\right]\label{Pf},\\
 \f\!\star\!\f(\xi,x,y,L)&\equiv&\int_{-i\infty}^{i\infty}\frac{dz}{2\pi i}\,\f\left(z-\xi/2\right)\f\left(z+\xi/2\right)\;\sin^2\frac{\pi(x-y-z+\xi/2)}{L},\label{conv2}\\
 \f\!\odot\!\f(\xi,x,y,L)&\equiv&\int_{-i\infty}^{i\infty}\frac{dz}{2\pi i}\,\f\left(z-\xi/2\right)\f\left(z+\xi/2\right)\;\frac12\sin\frac{2\pi(x-y-z)}{L}.\label{conv3}
 \ee
The second contribution in $\P_f$, controlled by $(f\!\odot\! f)$,  is a residue of a cancelation between the counter-term in the TT solution $\sim \int \frac12 f^2(z) c(z+y)$
and a corresponding term  from the
contraction of $j(z+y)$ in the TT solution and the exponential interaction. The latter gives rise to a delta function (canceling the TT counter-term) plus a remaining contribution, from which the second term in $\P_\f$ originates\footnote{ This can be seen by infinitesimally detaching the TT solution $\Phi$ from the left edge of the
exponential interaction.}.

The basic correlator for the zero momentum tachyon is given by the simpler expression
\be
\left\langle c\del c(x)\;\Phi(y)e^{-\int_{0}^{y}ds   J(s)}
\right\rangle_{C_{L}}=-\lambda^2\,\P_\f(x,y,L)\,e^{-\lambda^2 \G_\f(y,L)}.
\ee
The marginal and tachyon coefficients are finally given by the $\f$-dependent functionals
\be
\lambda_\SFT(\lambda)&=&\lambda \int_0^\infty d\ell\, \ell e^{-\ell}\int_0^1 d\hy \left(1+\lambda^2\,\F_\f(\ell+1/2,\ell \hy,\ell+1)\right)\,e^{-\lambda^2 \G_\f(\ell \hy,\ell +1)},\\
T(\lambda)&=&\frac{\lambda^2}{2}\int_0^\infty d\ell\,\ell(\ell+1)\, e^{-\ell}\int_0^1 d\hy\,
\P_\f(\ell+1/2,\ell \hy,\ell+1)\,e^{-\lambda^2 \G_\f(\ell \hy,\ell+1)},
\ee
where we introduced $\hy=y/\ell$ for later convenience.
Notice that the $\lambda$-dependence is fully manifest.
\subsection{Explicit results }

To continue further  we choose a family of functions $f_t(z)$, given
by the gaussians \cite{simple-marg}
\be
f_t(z)\equiv 2\lambda\sqrt{\pi}\,t\,e^{(tz)^2}.
\ee
As shown in \cite{simple-marg} the $t$ dependence is just an $L^-$ reparametrization of the TT solution $\Phi$ and it is thus a gauge redundancy. For very large $t$ the gaussian becomes a delta
function which localizes the exponential interaction to the boundary, providing  a regularization of contact term divergences, alternative to  the standard one by Recknagel and Schomerus \cite{bible}.
In our application this choice is particularly fortunate as it allows to perform the
convolution-like operations (\ref{conv1}, \ref{conv2}, \ref{conv3}) analytically.
In particular we have
\be
f_t*f_t(\xi)&=&\lambda^2\sqrt{2\pi}\,t\,e^{\frac{(t\xi)^2}{2}}\label{cconv1},\\
f_t\star f_t (\xi,x,y,L)&=&\lambda^2\sqrt{\frac{\pi}2}\,t\,e^{\frac{(t\xi)^2}{2}}\left(1-e^{\frac{\pi^2}{2L^2t^2}}\cos\left[\frac{2\pi(x-y+\frac\xi 2)}{L}\right]\right)\label{cconv2},\\
f_t\odot f_t (\xi,x,y,L)&=&\lambda^2\sqrt{\frac{\pi}2}\,t\,e^{\frac{(t\xi)^2}{2}}e^{\frac{\pi^2}{2L^2t^2}}\sin\frac{2\pi(x-y)}{L}\label{cconv3}.
\ee

The remaining integrations are performed numerically, except for the second term in $\P_f$ (\ref{Pf}) which can be computed analytically
\be
\int_{-i\infty}^{i\infty}\frac{d\xi}{2\pi i} \f\!\odot\!\f(\xi,x,y,L)\cos\frac{\pi\xi}{L}=\frac{\lambda^2}{2}e^{\frac{\pi^2}{L^2t^2}}\sin\frac{2\pi(x-y)}{L}.
\ee

\begin{figure}
 \begin{minipage}[b]{0.5\linewidth}
 \centering
 \resizebox{3.2in}{1.7in}{\includegraphics[scale=1]{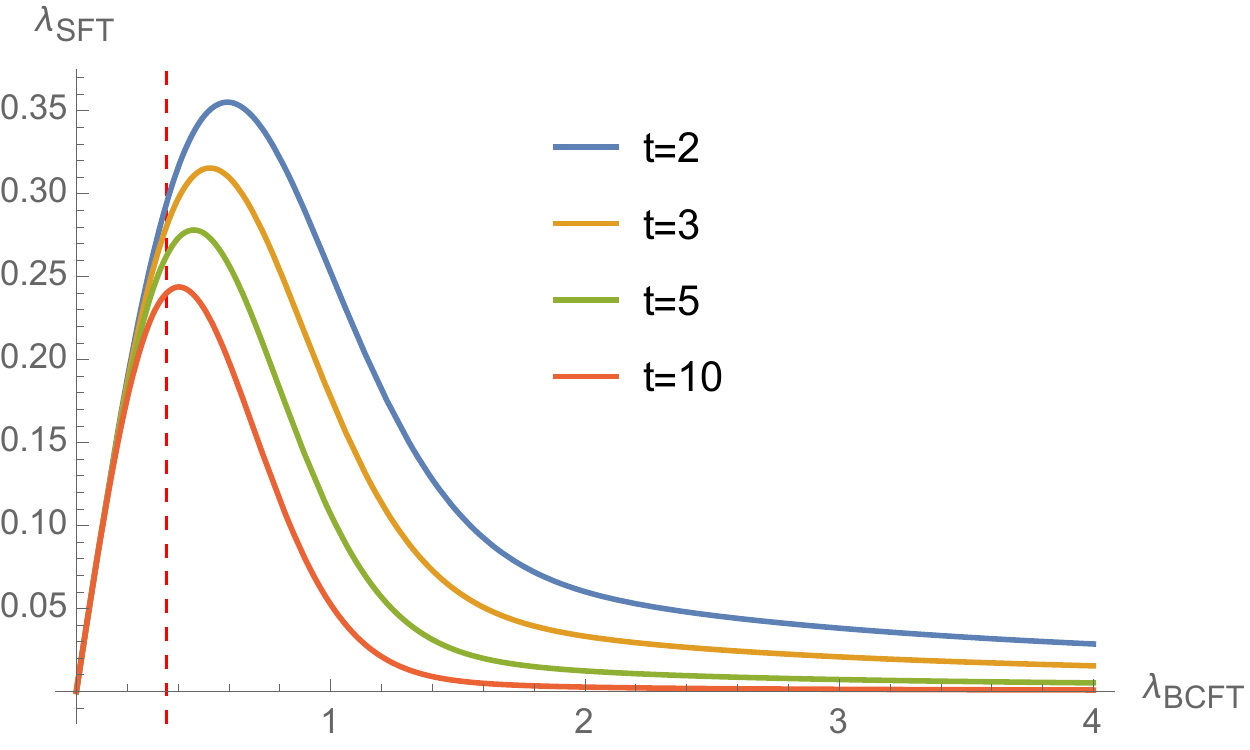}}
 \end{minipage}
 \hspace{0.5cm}
 \begin{minipage}[b]{0.5\linewidth}
 \centering
 \resizebox{3.2in}{1.7in}{\includegraphics[scale=1]{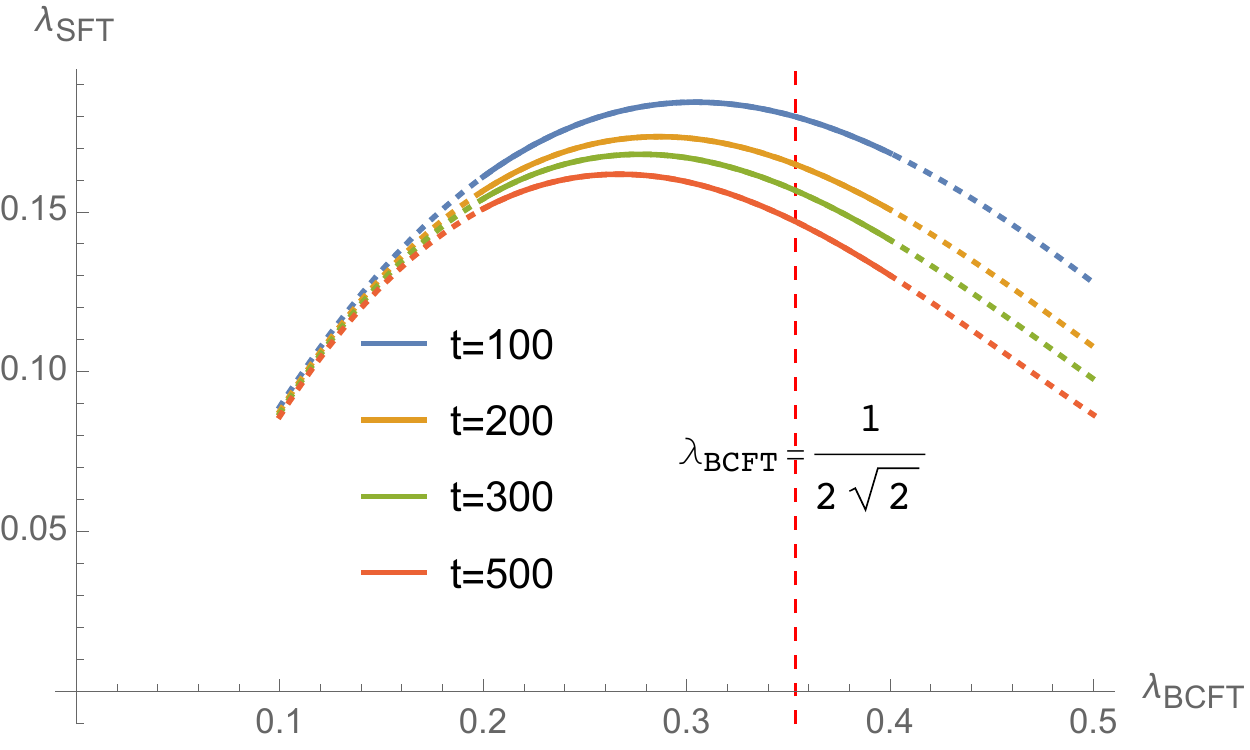}}
 \end{minipage}
 \caption{\label{fig:figure 2}{\small Plot of $\lambda_\SFT(\lambda)$ for a set of parameters $t$ in  the function $f_t(z)$ entering the definition of the TT solution:  $O(1)$ values of the gauge parameter $t$ are shown on the left  and larger values on the right. Dotted curves are extrapolations (for large $t$ the numerical integrations are very slow so we  selected a region around the peak). To set the scale: the   vertical dashed line corresponds to $\lambda_\BCFT=\frac1{2\sqrt2 }$ which, for the marginal deformation generated by $j(s)=\sqrt2\cos X(s)$, is the point where the initial Neumann boundary condition becomes Dirichlet. The  Siegel gauge solution stops existing approximately at  this point \cite{large2}. }}
 \end{figure}

In figure \ref{fig:figure 2} we plot the marginal coefficient as a function of $\lambda$ for a selection of $t$ parameters. The plots reveal a clear peak in $\lambda_\SFT$  and as a result for a given $\lambda_\SFT$ we find two corresponding values of $\lambda_\BCFT$. Had we included smaller values of $t$ (corresponding to less localized gaussians) we would have seen $\lambda_\SFT$ crossing the horizontal axis and approaching zero from below. This implies, in this smaller $t$ regime, a quadruple degeneracy for sufficiently small $\lambda_\SFT$, taking into account also negative values of $\lambda_\BCFT$. For $t\gtrsim 2$ the degeneracy is only two-fold and this is the region that we show in the plots. From the plots it is also quite evident that at large $\lambda$ the marginal field relaxes to zero.  Notice that the possibility that the maximum of $\lambda_\SFT$ is reached at the point $\lambda_\BCFT=\frac1{2\sqrt2}$ (which is  approximately what happens in Siegel gauge \cite{large2}, and for a range of $t$ parameters also here) is excluded to be true in general.  The position of the maximum is not gauge invariant.

\begin{figure}
 \begin{minipage}[b]{0.5\linewidth}
 \centering
 \resizebox{3.2in}{1.7in}{\includegraphics[scale=1]{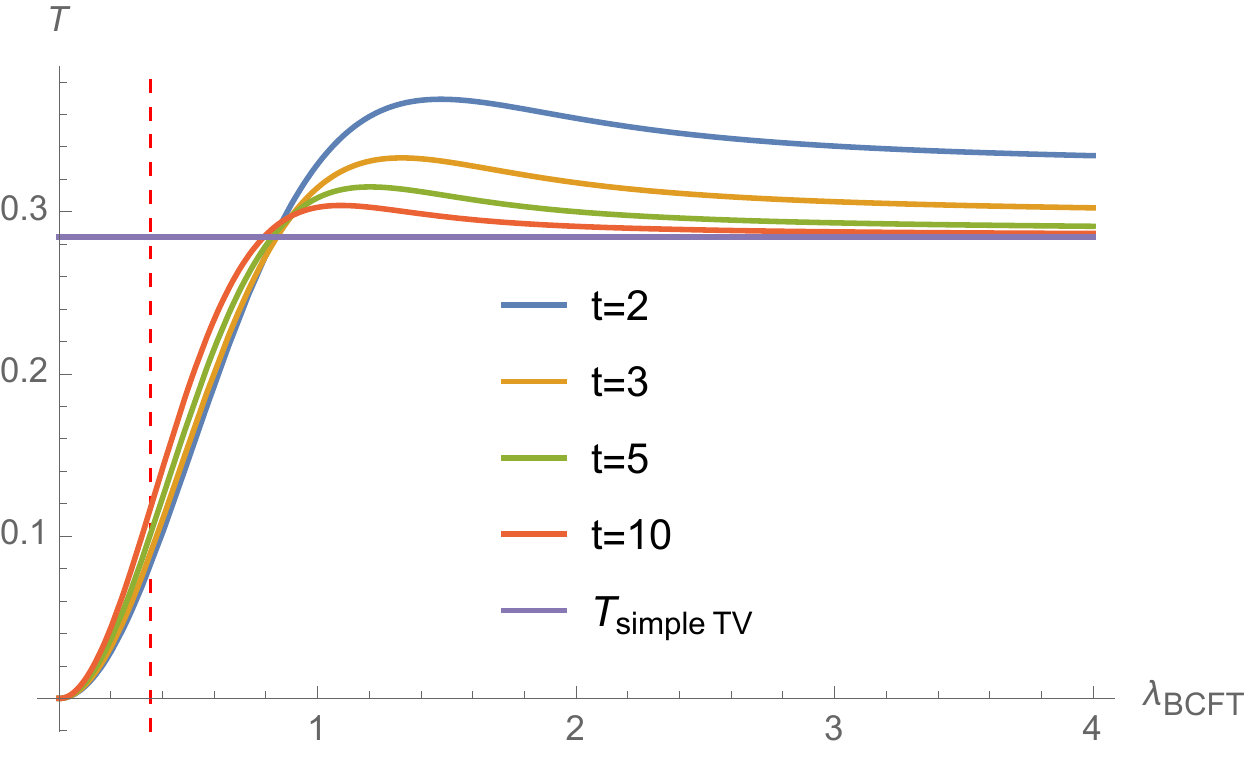}}
 \label{fig:figure1}
 \end{minipage}
 \hspace{0.5cm}
 \begin{minipage}[b]{0.5\linewidth}
 \centering
 \resizebox{3.2in}{1.7in}{\includegraphics[scale=1]{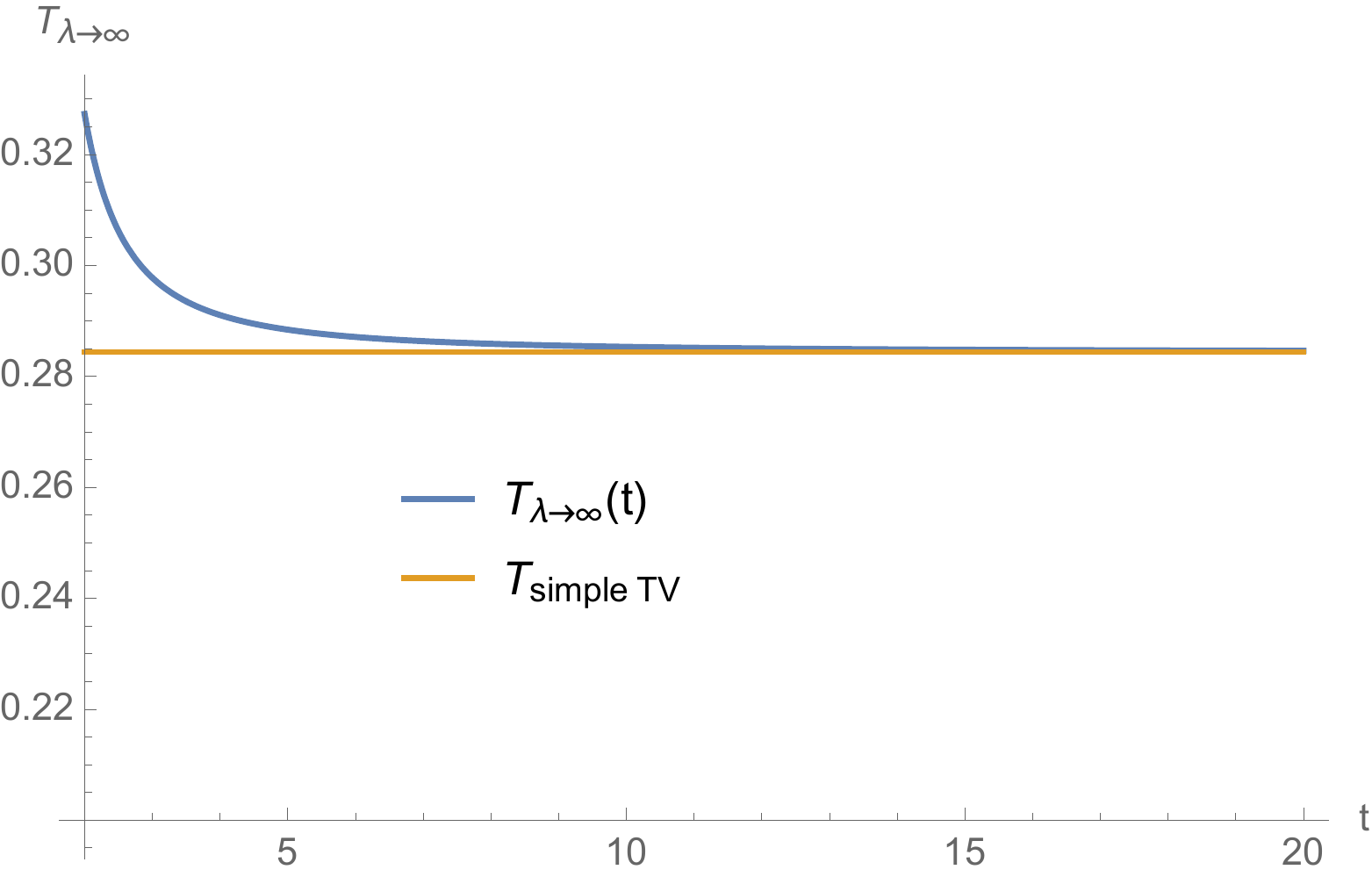}}
 \label{fig:figure3}
 \end{minipage}
 \caption{\label{Fig:large-tach}{\small On the left: Plot of the tachyon coefficient $T(\lambda)$ for a choice of $t$-parameters, together with the coefficient of the simple tachyon vacuum. The vertical dashed line corresponds to $\lambda_\BCFT=\frac1{2\sqrt2 }$. On the right: Exact large $\lambda$ asymptotic value for the tachyon coefficient as a function of the gauge parameter $t$; for large $t$ we recover the simple tachyon vacuum. }}
 \end{figure}

In figure \ref{Fig:large-tach} we plot the tachyon coefficient. Notice that for large $\lambda$ it tends to a positive constant. This positive constant, for large $t$, approaches the coefficient of the simple tachyon vacuum \mbox{$\Psi_{TV}=\frac1{1+K}c(1+K)Bc$} of \cite{simple}
\be\label{ES-coeff}
T_{\rm simple}=\frac1{4\pi}\int_0^\infty d\ell\,e^{-\ell}(\ell+1)^2\left(1-\cos\frac\pi{\ell+1}\right)=0.284394.
\ee

\subsection{Asymptotics for large $\lambda$ and large $t$ }
The numerical integrations we have just performed suggest that something non-trivial must happen  for large $\lambda$ since the marginal coefficient relaxes to zero, while the tachyon coefficient to a positive constant. Naively one would think that both quantities should relax to zero because of the exponential suppression $\sim e^{-\lambda^2 \G_\f}$, but evidently this is not the case. To understand what happens in the $\lambda\to\infty$ limit, we first notice that
\be
\lim_{\lambda\to\infty}\lambda^2 e^{-\lambda^2 \G_\f(\ell \hy,\ell+1)}=\delta\left[\G_\f(\ell \hy,\ell+1)\right].\label{local}
\ee
This is the leading term in the
asymptotic distributional expansion
\be
\lambda^2 e^{-\lambda^2 f(y)}=\sum_{n=0}^N (-1)^n\frac{\delta^{(n)}[f(y)]}{\lambda^{2n}}+O(\lambda^{-2N-2})\label{as-dis}.
\ee
Using the explicit definitions (\ref{G}, \ref{conv1}, \ref{cconv1}) this further simplifies to
\be
\lambda^2 e^{-\lambda^2 \G_\f(\ell \hy,\ell+1)}=\sqrt{\frac2\pi}     \frac{1}{\ell t}\delta(\hy)+O\left(\frac1{\lambda^2}\right).
\ee
This essentially means that the large $\lambda$ behavior is dominated by surfaces where the deformed region has zero width and the exponential term attains
 unit value

   $$e^{-\lambda^2 \G_\f(y\to 0,L)}=1.$$
Therefore the string field coefficients  for large $\lambda$ are not necessarily exponentially suppressed.
Let us start by looking at the fate of the tachyon coefficient. Using the above results we get
\be
\lim_{\lambda\to\infty}T(\lambda)=\frac1{2t}\sqrt{\frac2\pi}\int_0^\infty d\ell\,e^{-\ell}(\ell+1)\,\left[\lim_{\hy\to 0}\P_\f\left(\ell+1/2, \ell \hy, \ell+1\right)\right].
\ee
We can further compute
\be
\lim_{\hy\to 0}\P_\f\left(\ell+1/2, \ell \hy, \ell+1\right)=t\sqrt{\frac\pi2}\frac{(\ell+1)}{2\pi}\left(1-e^{\frac{\pi^2}{2(\ell+1)^2t^2}}\cos\frac\pi{\ell+1}\right).
\ee
The large $\lambda$ asymptotic value for the tachyon coefficient  is thus given by
\be
T(\lambda)=\frac1{4\pi}\int_0^\infty d\ell\,e^{-\ell}(\ell+1)^2\left(1-e^{\frac{\pi^2}{2(\ell+1)^2t^2}}\cos\frac\pi{\ell+1}\right)+O\left(\frac1{\lambda^2}\right),
\ee
and it is shown in figure \ref{Fig:large-tach}.

Notice that for very large $t$ this quickly approaches the tachyon coefficient (\ref{ES-coeff}) of the simple tachyon vacuum. On the contrary, as expected, the $t\to 0$ limit is very badly behaved which is related to the identity singularities of the TT solution \cite{simple-marg}.

If we apply the same analysis to $\lambda_{\rm SFT}$ we now find
\be
\lambda_{\rm SFT}(\lambda)= \sqrt{\frac{2}{\pi}}\frac\lambda t \int_0^{\infty}d\ell\,e^{-\ell}\left[\lim_{\hy\to 0}\S_\f\left(\ell+1/2, \ell \hy, \ell+1\right)\P_\f\left(\ell+1/2, \ell \hy, \ell+1\right)\right]+O\left(\frac1\lambda\right).
\ee
Now the $\hy\to0$ limit also includes $\S_\f$ (\ref{Sf}), and it is not difficult to see that the limit vanishes as it is the difference of two identical converging integrals. Therefore the $\lambda$-coefficient in the asymptotic expansion vanishes and we are left with\footnote{If needed, the precise $t$-dependent coefficient of $\lambda^{-1}$ can be computed
by taking into account one subleading correction in (\ref{as-dis}).}
\be
\lambda_{\rm SFT}(\lambda)=O\left(\frac1\lambda\right),\quad \lambda\to\infty,
\ee
which is indeed  much milder than the naively expected exponential suppression.

At last we would like to extract the $t\to\infty$ limit of our solution at fixed $\lambda$, where the exponential interaction $e^{-\int ds J(s)}$ localizes to the boundary. In order to do so we should note that the $\G$ function (\ref{G}), diverges  for large $t$ unless $\hy=0$ (which corresponds to a vanishing-width deformed region). To extract the relevant behavior in this limit it is useful to use the asymptotic formula
\be
\int_{-\infty}^{\infty}d\xi e^{-b\xi^2}\ln\left(1+\frac {a^2}{\sinh^2 \omega \xi}\right)\sim\frac{2\pi a}{\omega},\quad a\to0,
\ee
which allows to extract the small $\hy$ contribution from $\G$
\be
\G(\ell \hy,\ell+1)\sim \frac{(\ell+1)t}{\sqrt{2\pi}}\sin\frac{\pi   \hy \ell}{\ell+1},\quad \hy\to0.
\ee
Then, in the $t\to\infty$ limit we get the same localization mechanism (\ref{local}) as in the large $\lambda$ case, where now the  role of  large $\lambda^2$ is played by  large $t$, for fixed $\lambda$
\be
e^{-\lambda^2\G(\ell \hy,\ell+1)}\sim\sqrt{\frac2\pi}\frac1{\lambda^2 t \ell}\delta(\hy).
\ee
Following the same steps as for the $\lambda\to\infty$ case, we now find
\be
\lim_{t\to\infty}T(\lambda)&=& \left\lbrace \begin{array}{ll} \displaystyle\frac1{4\pi}\int_0^\infty d\ell\,e^{-\ell}(\ell+1)^2\left(1-\cos\frac\pi{\ell+1}\right)=T_{\rm simple},&\quad\lambda\neq0,\\ \\
0,&\quad\lambda=0,\end{array}\right.\\
\lim_{t\to\infty}\lambda_{\rm SFT}(\lambda)&=&0.
\ee

It is difficult to directly compare these limits with  the data because the numerical integrations are very slow in this region, but we have checked that the height and the position of the peaks in $\lambda_\SFT(\lambda,t)$  in figure \ref{fig:figure 2} are nicely fitted by
\be
\lambda_\SFT^{\rm max}(t)&\sim& 0.36\left(\frac1{\ln t}\right)^{0.44},\\
\lambda_\BCFT^{\rm crit}(t)&\sim& 0.59\left(\frac1{\ln t}\right)^{0.44},
\ee
which confirm our analysis for $t\to\infty$.
\section*{Acknowledgments}
\noindent

We thank Ted Erler, Mat\v{e}j Kudrna, Masaki Murata, Yuji Okawa and Barton Zwiebach for discussions. We thank the organizers of ``New frontiers in theoretical physics'', Cortona, May 2014 and ``String field theory and related aspects'', Trieste, July 2014, where our preliminary results were presented.
CM thanks the Academy of Science of Czech Republic for kind hospitality and support during part of this work.
The research of CM is funded by a {\it Rita
Levi Montalcini} grant from the Italian MIUR.
The research of MS has been supported by the Grant
Agency of the Czech Republic, under the grant 14-31689S.

\end{document}